\documentstyle{article}
\def\npag{\thanks{%\hspace* {-18pt}
This paper was done under the auspices of a CNCSIS Grant.}}
\date{\npag}

\title{ Hubble Red Shift and 
the Anomalous Acceleration of Pioneer 10 and 11}

\author{Kostadin Tren\v{c}evski$^{*}$\\
Faculty of Natural Sciences and Mathematics,\\ 
P.O.Box 162, 1000 Skopje, Macedonia}

\begin{document} 
\maketitle 

{\renewcommand{\thefootnote}{}%
\footnote{$^*$Electronic address: kostatre@iunona.pmf.ukim.edu.mk}%
\setcounter{footnote}{0}}%

%Keywords: Hubble constant, gravitational potential, fifth dimension 
PACS number: 98.80.Jk

\begin{abstract}
The basic relationship 
between the Hubble constant $H$ and the apparent anomalous acceleration 
$a_P$, which appears in the motion of the spacecraft Pioneer 10 and 11, 
is given by $a_P=cH$ \cite{1}. 
Using this equality, both Hubble red shift 
and the anomalous acceleration $a_P$ are explained, 
assuming that the gravitational potential in the universe changes linearly: 
$\Delta V=-c^2H\Delta t$. 
As a consequence after each second 
the time is faster $1+\frac{1}{3,76\times 10^{17}}$ times. 
%In the last section 
%the gravitational potential $V$ is considered as fifth dimension. 
\end{abstract}

%\section{Introduction} 

Hubble in 1928 has discovered that the velocities of the galaxies 
with respect to the Earth are proportional to the corresponding 
distances from the Earth. Namely, if $v$ is the velocity 
of a chosen star, and $R$ is its distance from the Earth, then $v=RH$,
where $H$ is an universal constant called Hubble constant. Some recent 
measurements show that $H$ is about 72 km/s/Mpc \cite{4}. 
We shall assume that it is about 82 km/s/Mpc 
$\approx (11,9\times 10^9 \hbox {years} )^{-1}\approx (3,76\times 10^{17}
\hbox {s})^{-1} \approx 2,66\times 10^{-18}$s$^{-1}$. It is 
accepted that the age of the universe is about $\frac{1}{H}$, and 
the universe is expanding. The velocities among the galaxies 
are sometimes up to $c/5$. These velocities 
are determined using the Doppler effect. 

%\section{Main results} 

In this paper it is shown that the previous Hubble law can be explained 
in another way, as a consequence of the change of the gravitational 
potential in the universe. Hence it will follow that such large 
velocities among the galaxies are only apparent, 
because the main effect is the red shift, 
and the red shift appears according to additional gravitational 
potential $V$ in the universe, which changes linearly (or almost 
linearly) with the time, i.e. 
$\frac{\partial V}{\partial t}\approx const$.
Indeed, it is detected red shift such that 
$\nu =\nu_0 \Bigl ( 1-\frac{v}{c}\Bigr )$, and replacing $v=HR$, indeed it 
is observed that 
\begin{equation}
\nu =\nu_0 \Bigl ( 1-\frac{RH}{c}\Bigr ).\label{1}
\end{equation}
On the other hand, if the potential changes linearly with the time, after 
time $t=R/c$ we have additional potential 
$\frac{R}{c}\frac{\partial V}{\partial t}$, 
and hence we have a shift such that 
\begin{equation}
\nu =\nu_0 \Bigl ( 1+\frac{1}{c^2}\frac{R}{c}\frac{\partial V}{\partial t}
\Bigr ).\label{2}
\end{equation}

{\em Remark.} 
Note that the gravitational potential $V$ is 
considered to be larger near gravitational bodies compared with the 
potential where there is no gravitation. For example, 
near the spherical body we accept that $V=\frac{GM}{r}$. 
If we accept that 
$V=-\frac{GM}{r}$, then the sign "+" in the formula (\ref{2}) should be 
replaced by "-". 

Comparing the formulas (\ref{1}) and (\ref{2}) we obtain 
\begin{equation}
H=-\frac{1}{c^2}\frac{\partial V}{\partial t}.\label{3}
\end{equation}

Since $\frac{\partial V}{\partial t}<0$, 
in the past the time in the universe 
was slower than now, and in future it will be faster than now. 
Indeed, each second the time is faster approximately 
$\lambda = (1+H\times 1$s$)=1+\frac{1}{3,76\times 10^{17}}$ times. 

According to (\ref{3}) we obtain that 
\begin{equation}
\frac{\partial V}{\partial t}\approx -2,4\times 10^3
\frac{\hbox {cm}^2}{\hbox {s}^3}.\label{4}
\end{equation}
This shows that the potential energy of mass of 1kg arises for 
$2,4\times 10^3 
\frac{\hbox {kg}\cdot \hbox{cm}^2}{\hbox {s}^2}=0,24 \hbox {J}$ per 
second. Probably this energy comes from 
the dark energy which is about 67\% in the universe \cite{4}. 

On the other hand, in the last decades are considered the motions 
of the spacecraft Pioneer 10 and 11, by comparing the initial 
frequency of a signal which is sent from the Earth to the spacecraft 
and the frequency of the signal which comes back. 
The frequency of the 
received signal on the Earth does not fit with the predicted frequency 
modeled by using the Doppler effect, the position and motion of the 
spacecraft using the general relativity and a lot of perturbations 
(see Ref. \cite{1,2}). 
In the case of the spacecraft Pioneer 10 and 11 
we have a similar situation like with the red shift from the galaxies, 
but much more complicated. 
The observations show that it appears almost a constant apparent unmodeled 
acceleration 
$a_P\approx (8,74\pm 1,33)\times 10^{-8}$cm/s$^2$, 
which seems to act to the spacecraft toward the Sun. 
The acceleration $a_P$ is introduced by formula (15) in \cite{1}, i.e. 
\begin{equation}
[\nu_{obs}(t)-\nu_{model}(t)]_{DSN} =-\nu_0\frac{2a_Pt}{c},
\qquad \nu_{model}=
\nu_0\Bigl [1-2\frac{v_{model}(t)}{c}\Bigr ],\label{5}
\end{equation}
where $\nu_0$ is the reference frequency, the factor 2 is because of two way 
data, $v_{model}$ is the modeled velocity of the spacecraft due to the 
gravitational and other forces, and $\nu_{obs}$ is frequency of the 
re-transmitted signal observed by DSN antennae. 
After time $2t$ 
has been detected a small blue shift on the top of the 
red shift caused by the motion of the spacecraft 
outwards the Sun. {Form} (\ref{5}), an unexplained blue shift 
\begin{equation}
\nu = \nu_0\Bigl (1+\frac{2a_Pt}{c}\Bigr ),\label{6}
\end{equation}
is detected, where $2t$ is the time of the light signal in two directions. 
In \cite{1} it is mentioned that without using 
the acceleration $a_P$, the anomalous frequency shift can be 
interpreted by 
"clock acceleration" $-a_t=-2,8\times 10^{-18}$s/s$^2$. 
This causes just the blue shift given by (\ref{6}). 
This model assumes that the 
time is uniform, but there is only technical problem with the 
clocks. Although this model gives good explanations for both frequency shift 
and the trajectories, it is rejected \cite{1}. 
Further we shall use this 
model for comparison with our model via the time dependent potential. 

Many people have noticed that 
the acceleration $a_P$ and the Hubble 
constant $H$ are related by (see Ref. \cite{1})
\begin{equation}
a_P=cH,\label{7}
\end{equation}
and some possible explanations about the anomalous acceleration were 
given (for example \cite{3,5,6,7}). 
Indeed, if we assume that $H$=82km/s/Mpc, then from (\ref{7}) for $a_P$ we 
obtain 
$8\times 10^{-8}$cm/s$^2$. For this reason we assumed that 
$H$=82km/s/Mpc. 

Now we shall explain the apparent acceleration $a_P$ 
connected with the frequency shift. 
Let us denote by $X,Y,Z,T$ our natural coordinate system, in 
the deformed space-time, and 
let us denote by $x,y,z,t$ the normed coordinates of an imagine 
coordinate system, where the space-time is "uniform", except close to the 
gravitational objects. 
Then, according to the 
general relativity we have the following equalities 
$$
dx=\Bigl (1+\frac{V}{c^2}\Bigr )^{-1}dX 
= \Bigl (1+tH\Bigr )dX,\qquad 
dy=\Bigl (1+\frac{V}{c^2}\Bigr )^{-1}dY = \Bigl (1+tH\Bigr )dY,
$$
\begin{equation}
dz=\Bigl (1+\frac{V}{c^2}\Bigr )^{-1}dZ 
= \Bigl (1+tH\Bigr )dZ,\qquad 
dt=\Bigl (1-\frac{V}{c^2}\Bigr )^{-1}dT = \Bigl (1-tH\Bigr )dT.
\label{8}
\end{equation}

{\em Remark.} Note that it is sufficient in this paper to assume that 
(8) are satisfied. Then it is not necessary to speak about the 
time dependent gravitational potential. The coefficients 
$1+tH$ and $1-tH$ in (\ref{8}) should be exponential functions, but 
neglecting $H^2$ and smaller quantities we accept these linear functions. 

{From} (\ref{8}) we obtain 
$$
\Bigl (\frac{dX}{dT},\frac{dY}{dT},\frac{dZ}{dT}\Bigr )=
\Bigl (\frac{dx}{dt},\frac{dy}{dt},\frac{dz}{dt}\Bigr )(1-2tH)
$$
and by differentiating this equality by $T$ we get 
$$
\Bigl (\frac{d^2X}{dT^2},\frac{d^2Y}{dT^2},\frac{d^2Z}{dT^2}\Bigr )=$$
$$=\Bigl (\frac{d^2x}{dt^2},\frac{d^2y}{dt^2},\frac{d^2z}{dt^2}\Bigr ) 
-3tH \Bigl (\frac{d^2x}{dt^2},\frac{d^2y}{dt^2},\frac{d^2z}{dt^2}\Bigr )
-2H\Bigl (\frac{dX}{dT},\frac{dY}{dT},\frac{dZ}{dT}\Bigr ).
$$
In normed coordinates $x,y,z,t$ there is no acceleration caused by the 
time dependent gravitational potential. Thus, 
in real coordinates $(X,Y,Z,T)$ appears an additional acceleration 
$$
-3tH \Bigl (\frac{d^2x}{dt^2},\frac{d^2y}{dt^2},\frac{d^2z}{dt^2}\Bigr )
-2\Bigl (H\frac{dX}{dT},H\frac{dY}{dT},H\frac{dZ}{dT}\Bigr ).
$$
The first component 
$-3tH \Bigl (\frac{d^2x}{dt^2},\frac{d^2y}{dt^2},\frac{d^2z}{dt^2}\Bigr )$
is smaller than the second component 
\begin{equation}
-2\Bigl (H\frac{dX}{dT},H\frac{dY}{dT},H\frac{dZ}{dT}\Bigr ) = 
-2\Bigl (\frac{a_P}{c}\frac{dX}{dT},
\frac{a_P}{c}\frac{dY}{dT},\frac{a_P}{c}\frac{dZ}{dT}\Bigr ),\label{9}
\end{equation}
and so it is of minor role in the explanation of the frequency shift of 
the Pioneer spacecraft. Indeed, the first acceleration is also important, 
but not so much in the case of the spacecraft. 
Further, in our simplified model will be used only the acceleration (9). 

We shall explain why it is measured the blue shift given by (\ref{6}), 
instead of the red shift like in (\ref{1}). 

For the sake of simplicity we assume that the spacecraft is moving 
radially in the solar system outwards the Sun, and the DSN antenna
has a fixed position in the solar system, 
collinear with the Sun and the spacecraft. 
Assume that far from the Sun, when the spacecraft is on distance $R$
from the DSN antenna, 
its Newtonian acceleration is approximately a constant acceleration 
equal to $-a$, 
i.e. $a$ towards the Sun. 
When it is on distance $R$, for $t=0$, assume that its velocity is $v_0$.
Without loss of generality we can deal according to the Newton theory, 
when it is possible. 
Let us calculate 
$\frac{\Delta \nu}{\nu_0}=(\nu_{obs}-\nu_0)/\nu_0$, neglecting the terms 
containing $H^2$. 

If $v_0=0$ and $a=0$, then analogous to (\ref{1}), after time $2R/c$ in two 
directions we have red shift 
$$
\frac{\Delta \nu}{\nu_0}=-2H\frac{R}{c}.
$$
Since the real acceleration is equal to $-a-2Hv$, it satisfies the 
differential equation $\frac{dv}{dt}=-a-2Hv$ and hence 
$\frac{d^2v}{dt^2}=-2H\frac{dv}{dt}\approx 2Ha$. Thus, using the Taylor's 
series for the velocity $v$ we get 
$$
v=v_0+(-a-2Hv_0)t+aHt^2.
$$
Then for the observed shift according to our model we obtain
$$
\Bigl (\frac{\Delta \nu}{\nu_0} \Bigr )_I = 
-2H\frac{R+v_0t-\frac{1}{2}at^2}{c}
-2\frac{v_0+(-a-2Hv_0)t+aHt^2}{c},
$$ 
$$
\Bigl (\frac{\Delta \nu}{\nu_0} \Bigr )_I = 
2H\frac{R+v_0t-\frac{1}{2}at^2}{c}
-2\frac{v_0-at}{c}-4H\frac{R}{c}.
$$ 
On the other side, if we neglect the acceleration (\ref{9}), 
and consider the blue
shift (\ref{6}), i.e. if we use "clock acceleration" model, then 
$$\Bigl (\frac{\Delta \nu}{\nu_0}\Bigr )_{II} = 
2H\frac{R+v_0t-\frac{1}{2}at^2}{c}
-2\frac{v_0-at}{c}.$$ 
Comparing the frequencies in both cases $\nu_I$ and $\nu_{II}$, we see that 
$$\frac{d}{dt}\Bigl (\frac{\Delta \nu}{\nu_0}\Bigr )_I=\frac{d}{dt}
\Bigl (\frac{\Delta \nu}{\nu_0}\Bigr )_{II},
\quad \hbox {i.e.} \quad \frac{d\nu_I}{dt}=\frac{d\nu_{II}}{dt}.$$
Although $\nu_I\neq\nu_{II}$, it is more important that their derivatives 
are equal, because the determination of $a_P$ can not be done by a 
single measurement, but statistically followed on long time intervals. 
This explains the blue shift given by (\ref{6}), according to our model. 

Appart from the previous explanation, we shall give now an explicit formula 
for the anomalous acceleration $a_P$ in general case. 

Let $R=R(T)$ be the distance from the spacecraft to the DSN antenna. Let us 
calculate $\frac{\Delta \nu}{\nu_0}=(\nu_{obs}-\nu_0)/\nu_0$. 
According to our model, neglecting the terms containing $H^2$, 
this expression is equal to 
$$\frac{\Delta \nu}{\nu_0} = -2H\frac{R}{c}-2\frac{\frac{dR}{dT}}{c},$$
where the expression $-2H\frac{R}{c}$ corresponds to the Hubble red shift 
from (\ref{2}). 
On the other side, according to (\ref{5}), where $v_{model}$ it determined 
without using the Hubble constant $H$, we have 
$$\frac{\Delta \nu}{\nu_0} = 
2a_P\frac{R}{c^2}-2\frac{(\frac{dR}{dT})_{H=0}}{c}.$$ 
{From} the last two equalities we are able to find 
the expression for $a_P$ 
$$a_P=-c\Bigl (\frac{1}{R}\frac{dR}{dT}-\frac{1}{R}
\Bigl (\frac{dR}{dT}\Bigr )_{H=0}+H\Bigr ).$$
Using that 
$\frac{dR}{dT}=\frac{dr}{dt}(1-2HT)=
\Bigl (\frac{dR}{dT}\Bigr )_{H=0}(1-2HT)$ according to (\ref{8}), 
we obtain finally 
\begin{equation}
a_P=cH\Bigl (2\frac{dR}{dT}\frac{T}{R}-1\Bigr ).\label{10}
\end{equation}
This equality gives the required expression of the 
acceleration $a_P$. Now we are not able to determine the initial 
value of $T$, i.e. when we should start to measure the time $T$ in 
(\ref{10}). It depends on 
the estimation of the initial values for the motion of the spacecraft, i.e. 
it depends on the departure of the real initial values. Thus, for different 
spacecraft are obtained close but different values of $a_P$. Note also that 
for each initial value of $T$, when $T$ tends to infinity (or $R$ tends to 
infinity), then $a_P$ tends to $cH$. 
In an ideal case, if the experiment is done in an 
inertial system, then $dR/dT=R/T$ is a constant velocity and then 
$a_P\equiv cH$.

The change of the gravitational potential 
causes slight changes in the results of the known 
experiments about general relativity, but they are negligible. 
The change of the results in the test of 
Shapiro time delay is negligible if we consider radio 
signals on short distances like in the solar system. 
Indeed, the change of the potential in the universe is about 
$2,4\times 10^3$cm$^2$/s$^2$ each second, while the gravitational 
potential on 1AU from the Sun is equal to $30^2$km$^2$/s$^2=
9\times 10^{12}$cm$^2$/s$^2$. This potential difference will be achieved 
from the universe after 118 years, which is too long compared with the 
time needed for an experiment about Shapiro time delay. The change of the 
deflection of the light rays near the Sun is completely negligible up to 
$c^{-2}$. The angle between two perihelions is changed additionally of 
order $\frac{\Theta^2 H^2}{\epsilon}$, 
where $\Theta$ is the orbit period of the planet 
and $\epsilon$ is the eccentricity. 
This change of the angle is negligible with respect to the measured angle 
$\frac{6GM\pi }{ac^2(1-\epsilon^2)}$. 

%Note that the normed coordinates $x,y,z,t$ from (\ref{8}) 
%can not solve the problem of motion in gravitational field 
%in general case, like it was done for the special choice of the 
%potential $V$. Indeed, the gravitational potential 
%also changes in the normed 
%coordinates $x,y,z,t$, but the corresponding acceleration (\ref{9}) 
%disappears as a gradient of the potential. 

\end{document}